\documentclass[12pt]{article}

\usepackage{amsmath,amssymb,amsthm,mathtools}
\usepackage{setspace,enumitem,soul,url,xcolor,hyperref,bm}
\usepackage{tikz-cd}

\newtheorem{example}{Example}

\usepackage[T1]{fontenc}
\usepackage{tgbonum} \fontfamily{qbk} 

\begin{document}

\title{Sampling for network function learning}
\author{\emph{Li-Chun Zhang}}
\date{}
\maketitle

Given a \emph{valued graph}, where both the nodes and the edges of the graph are associated with one or several values, any network function for a given node must be defined in terms of that node \emph{and} its connected nodes in the graph. Generally, applying the same definition to the whole graph or any given subgraph of it would result in systematically different network functions. In this paper we consider the feasibility of graph sampling approach to network function learning, as well as the corresponding learning methods based on the sample graphs. This can be useful either when the edges are unknown to start with or the graph is too large (or dynamic) to be processed entirely. 


\section{Network function}

Let $G = (U,A)$ be a simple graph with $U$ as the set of nodes, $N = |U|$, and $A$ as the adjacency matrix, where $a_{ij} = 1$ if an edge from $i$ to $j$ exists and 0 otherwise. By definition, an undirected graph is such that $a_{ij} \equiv a_{ji}$ for any $i,j\in U$ and, as a convention for this paper, we assume $a_{ii} \equiv 0$ for any $i\in U$, i.e. there are no loops in the graph.
Let $v_G$ contain the values associated with the graph, which are $y_U = \{ y_i : i\in U\}$, $x_U = \{ x_i : i\in U\}$ and $\omega_A = \{ \omega_{ij} : i,j \in U, a_{ij} =1 \}$. Whereas $y_i$ is a scalar \emph{outcome} value the function targets, $x_i$ and $\omega_{ij}$ are (possibly vectors of) \emph{feature} values that can be used to form the arguments of the function. 

Given the \emph{valued graph} $(G, v_G)$, a \emph{network function (NF)} for any $i$ in $U$ must be defined in terms of $i$ \emph{and} its \emph{(graph) neighbours}, which are given by
\[
\nu_i = \{ j\in U : a_{ij} + a_{ji} > 0\}
\] 
For any $i\in U$, let its degree, out-degree and in-degree be
\[
d_i = |\nu_i |\qquad d_{i+} = \sum_{j\in \nu_i} a_{ij} \quad\text{and}\quad d_{+i} = \sum_{j\in \nu_i} a_{ji} 
\]
where $d_i = d_{i+} = d_{+i}$ in undirected graphs by definition.

\subsection{Contextual network function (CNF)}

For any $i\in U$, let $x_{\nu_i} = \{ x_j : j\in \nu_i\}$ and $\omega_{\nu_i} = \{ \omega_{ij}, \omega_{ji} : j\in \nu_i\}$. A \emph{contextual NF} explicates a mapping $(x_i, x_{\nu_i}, \omega_{\nu_i}) \mapsto \mu_i \in \mathbb{R}$, defined by \begin{equation} \label{CNF} 
\mu_i = f(x_i, \dot{x}_i; \theta_0) \quad\text{and}\quad \theta_0 = \arg \min_{\theta} D\big( \mu_U(\theta), y_U \big) 
\end{equation}
given \emph{contextual features}  
\begin{equation} \label{dot}
\dot{x}_i = \eta(x_{\nu_i}, \omega_{\nu_i}) 
\end{equation}
and the \emph{parameter} value $\theta_0$ that minimises a metric $D$ of the distance between $y_U$ and $\mu_U(\theta) = \{ \mu_i(\theta) : i\in U\}$. 

For instance, a simple model of a random outcome $y_i$ can be given as
\begin{align*} 
y_i & = x_i^{\top} \beta + z_i^{\top} \gamma + e_i \\
z_i & = d_{+i}^{-1} \sum_{j\in \nu_i} a_{ji} x_j 
\end{align*}
where $z_i$ is the average feature of those neighbours with $a_{ji} =1$, and $e_i$ is an independent mean-0 random error, and $(\beta, \gamma)$ are unknown constants. The corresponding contextual NF with $\theta = (\beta, \gamma)$ is 
\[
\mu_i = E(y_i \mid x_i, x_{\nu_i}, \omega_{\nu_i} ; \beta, \gamma) 
= x_i^{\top} \beta + z_i^{\top} \gamma
\]  
As Friedkin (1990) comments, a group average effect via $z_i$ may be unrealistic in many social settings, where the nodes represent a finite population of persons and the edges the existent influences among them. 

Potentially more realistic contextual features can be engineered in terms of the \emph{neighbour function} \eqref{dot}, which must be invariant over permutations of the adjacent nodes. For instance, varying strength of network influence can be achieved via $\omega_{ji}$ associated with $(ji)\in A$. Finally, note that the definition \eqref{CNF} does not require any distributional assumptions about $y_U$, which are treated as unknown constants and the targets of $\mu_U$.

\subsection{Recursive network function (RNF)}

A \emph{recursive NF} defines a mapping $(x_i, \mu_{\nu_i}) \mapsto \mu_i$, where $\mu_{\nu_i} = \{ \mu_j : j\in \nu_i\}$, such that $\mu_i$ may depend on some $\mu_j$ for $j\in \nu_i$, and $\mu_j$ itself may depend on some $\mu_k$ for $k\in \nu_j$, and so on recursively. We consider any recursive NF defined by
\begin{equation} \label{RNF}
\mu_i = \lambda_0 \dot{\mu}_i + c_i \quad\text{and}\quad \theta_0 = \arg \min_{\theta} D\big( \mu_U(\theta), y_U \big) 
\end{equation}
where $\theta = (\beta, \lambda)$, given $c_i = c(x_i; \beta_0)$ and any \emph{linear} neighbour function  
\begin{equation} \label{lineardot}
\dot{\mu}_i = \sum_{j\in \nu_i} m_{ij} \mu_j
\end{equation}
where $m_{ij}$ depends on $\omega_{\nu_i}$, and $\lambda_0$ is subject to the restriction 
\[
\| \lambda_0 M \| < 1
\]
where the matrix $M$ has $m_{ij}$ as its  $(i,j)$th element if $j\in \nu_i$ and 0 otherwise. 

Given linear neighbour function \eqref{lineardot}, we can rewrite \eqref{RNF} in matrix notation as
\[
\mu = \lambda_0 M \mu + c
\] 
The restriction $\| \lambda_0 M \| < 1$ ensures that $Q$ exists for 
\[
\mu = Q c \coloneqq (I - \lambda_0 M)^{-1} c
\]
where $I$ is the $N\times N$ identity matrix. Thus, the recursive NF defines a mapping $(x_U, \omega_A) \mapsto \mu_i$, which generally depends on the entire adjacent matrix $A$. 

Meanwhile, let the $\tau$th order approximation to the matrix $Q$ be given as
\[
Q_{\tau} \coloneqq I + \lambda_0 M + \lambda_0^2 M^2 + \cdots + \lambda_0^{\tau} M^{\tau}
\]
Since $M$ has the same non-zero elements as $A^{\top}$, the $(i,j)$th element of $M^t$ depends on all the possible paths of length $t$ from some node $j$ to $i$. Let $\alpha_i^t$ contain the nodes that can reach $i$ in exactly $t$ steps, which correspond to the non-zero elements in the $i$th column of $A^t$. Then, by the approximation $\mu \doteq Q_{\tau} c$, $\mu_i$ would only depend on $c_j$ if $j$ belongs to the $\tau$th-order neighbourhood 
\begin{equation} \label{ext}
\nu_i^{\tau}= \bigcup_{t=1}^{\tau} \alpha_i^t 
\end{equation}
This shows how the recursive NF can represent a power series of contextual effects (or influences) over any number of steps.

\begin{example}
Take the digraph in Figure \ref{fig:toy} for an illustration. 
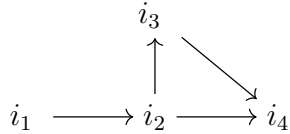
\begin{figure}[ht] 
\centering
\begin{tikzcd}[cramped]
& i_3 \arrow[dr] & \\  
i_1 \arrow[r] & i_2 \arrow[r] \arrow[u] & i_4 
\end{tikzcd} 
\caption{A digraph with 4 nodes} \label{fig:toy}
\end{figure} \\
Arrange the rows of the adjacency matrix $A$ in the order $i_1, i_2, i_3, i_4$. We have
\[
A = \left[ \begin{array}{cccc} 0 & 1 & 0 & 0\\ 0 & 0 & 1 & 1\\ 0 & 0 & 0 & 1\\ 0 & 0 & 0 & 0 \end{array} \right] \qquad
A^2 = \left[ \begin{array}{cccc} 0 & 0 & 1 & 1\\ 0 & 0 & 0 & 1\\ 0 & 0 & 0 & 0\\ 0 & 0 & 0 & 0 \end{array} \right] \qquad
A^3 = \left[ \begin{array}{cccc} 0 & 0 & 0 & 1\\ 0 & 0 & 0 & 0\\ 0 & 0 & 0 & 0\\ 0 & 0 & 0 & 0 \end{array} \right] 
\]
For the 2nd-order approximation to the RNF, $\mu_{i_1}$ does not depend on any other $c_j$ since $d_{+i_1} =0$, and $\mu_{i_2}$ depends on $c_{i_1}$ where $\nu_{i_2}^2 =\alpha_{i_2}^1 = \{ i_1\}$, and $\mu_{i_3}$ depends on $c_j$ for $j\in \nu_{i_3}^2 = \{ i_1, i_2 \}$ where $\alpha_{i_3}^1 = \{ i_2\}$ and $\alpha_{i_3}^2 = \{ i_1\}$, and $\mu_{i_4}$ depends on all the other $c_j$ since $\nu_{i_4}^2 = \{ i_1, i_2, i_3 \}$. Note that $c_{i_2}$ affects $\mu_{i_4}$ via both $M$ and $M^2$, due to the edge $(i_2 i_4)$ and the 2-path $(i_2, i_3, i_4)$, respectively. Since $A^t = 0$ for $t\geq 4$, we have $Q = Q_{\tau}$ for any $\tau \geq 3$ given $\| \lambda_0 M \| < 1$.
\end{example}

\subsection{Remarks}

Ord (1975) considers maximum likelihood estimation of the model 
\[
y_i = \rho \sum_{j\in \nu_i} m_{ij} y_j + c_i + \epsilon_i \quad\text{and}\quad c_i = x_i^{\top} \beta 
\quad\text{and}\quad \sum_{j\in \nu_i} m_{ij} = 1
\] 
that allow interactions among locations $U$. This corresponds to the RNF 
\[
\mu_i = E(y_i \mid G, x_U, \omega_A; \beta, \rho) = \rho \sum_{j\in \nu_i} m_{ij} \mu_j + c_i
\]
However, there are several obstacles to applying Ord's approach to recursive NF learning generally. First, it requires known $\nu_i$ and $\{ m_{ij} : j\in \nu_i\}$ for all $i\in U$, not just a sample of them. Next, an assumption of normally distributed $\epsilon_i$ is needed, given which the log-likelihood requires one to compute 
\[
\mbox{det}(I - \rho M) = \prod_{i=1}^N (1 - \rho \lambda_i)
\]
where $\lambda_1, ..., \lambda_N$ are the eigenvalues of $W$, which can be too demanding if $N$ is very large even when $M$ is known. Finally, the assumption of independent $\epsilon_i$ with constant variance may be unrealistic. For instance, suppose $a_{ij} = a_{ji} =1$ and $i,j$ are not connected to any other nodes. Then, we would have
\[
\begin{cases} \mu_i = \rho \mu_j + c_i \\  \mu_j = \rho \mu_i + c_j \end{cases} \& ~~
\begin{cases} y_i = \mu_i + e_i \\  y_j = \mu_j + e_j \end{cases} \Rightarrow~~ 
\begin{cases} y_i = \rho y_j + c_i + (e_i -  \rho e_j) \\  y_j = \rho y_i + c_j + (e_j -  \rho e_i) \end{cases} 
\]

An alternative network disturbance model is sometimes adopted to account for node interdependencies (e.g. Ord, 1975; Leenders, 2002), where 
\[
y_i = x_i^{\top} \beta + e_i \quad\text{and}\quad e_i = \rho \sum_{j\in \nu_i} m_{ij} e_j + \epsilon_i
\] 
or
\[
y = X\beta + (I - \rho M)^{-1} \epsilon \quad\text{and}\quad e = \rho M e + \epsilon
\]
in matrix notation, which however corresponds to the non-network function $\mu_i = E(y_i | x_i) = x_i^{\top} \beta$ that is not our concern in this paper.

\section{Learning}

Let $e_U = \{ e_i : i\in U\}$ be the discrepancies between $\mu_U(\theta)$ and $y_U$, where $e_i = y_i - \mu_i$. Let $e_{\nu_i} = \{ e_j : j\in \nu_i\}$. We consider in this paper metrics $D$ of the form 
\[
D(\mu_U, y_U) = \sum_{i\in U} D(e_i, e_{\nu_i})
\]
such that given the population graph $G$ and all the associated values $v_G$, the parameter $\theta_0$ by definition \eqref{CNF} or \eqref{RNF} must be a solution to 
\[
\frac{\partial}{\partial \theta} D(\mu_U, y_U) = H(\theta; G) = \sum_{i\in U} H_i(\theta) 
= \sum_{i\in U} \frac{\partial}{\partial \theta} D(e_i, e_{\nu_i}) = 0
\]

Now, for NF learning we would estimate $\theta_0$ based on a sample of nodes and edges from $G$, to be referred to as the \emph{sample graph} and denoted by $G_s = (U_s, A_s)$, where $U_s$ contains the observed nodes of $U$ and $A_s$ the observed elements of the adjacency matrix $A$. Specifically, we shall estimate $\theta_0$ by solving the \emph{sample estimating equation (SEE)}
\begin{equation} \label{SEE}
H(\theta; G_s) = \sum_{i\in U_s} w_i H_i(\theta) = 0 
\end{equation}
where $w_i$ is an appropriately chosen weight for each $i\in U_s$, depending on the sampling method by which $(G_s, v_{G_S})$ are observed from $(G, v_G)$, such that the SEE is \emph{unbiased} under hypothetically repeated sampling, denoted by
\[
E\big[ H(\theta; G_s) \big] = \sum_{i\in U} H_i(\theta) = H(\theta; G)
\]
Note that it is inappropriate generally to let $w_i \equiv 1$ given we observe all the values from $v_G$ which are required for $H_i(\theta)$ and 0 otherwise, since
\[
E\big[ \sum_{i\in U_s : w_i =1} H_i(\theta) \big] = E\big[ \sum_{i\in U} \mathbb{I}(w_i =1) H_i(\theta) \big]  \neq  \sum_{i\in U} H_i(\theta) 
\]

Below we first propose the choices of $D$ and the corresponding SEEs \eqref{SEE} for CNF and RNF, respectively. We then explain how the weights in \eqref{SEE} can be obtained for two basic graph sampling methods, which yield either depth-first or breadth-first observation in graphs.

\subsection{SEE}

First, a natural choice of $D$ for the CNF \eqref{CNF} is 
\begin{equation} \label{D}
D(\mu_U, y_U) = \sum_{i\in U} \frac{e_i^2 }{2 \sigma_i^2} 
\end{equation}
where $\sigma_i^2$ is a chosen constant that may depend on $x_i$. The SEE follows as
\[
H(\theta; G_s) = \sum_{i\in U_s} \frac{w_i}{\sigma_i^2} \big( \frac{\partial \mu_i}{\partial \theta} \big) (\mu_i - y_i) = 0
\]
The similarity to weighted least squares for standard linear regression becomes clear given $\mu_i = u_i^{\top} \theta$, where $\theta^{\top} = (\beta^{\top}, \gamma^{\top})$ and $u_i^{\top} = (x_i^{\top}, z_i^{\top})$, such that
\[
\theta_0 = \big( \sum_{i\in U} \sigma_i^{-2} u_i u_i^{\top} \big)^{-1} \sum_{i\in U} \sigma_i^{-2} u_i y_i 
\]
and the SEE yields
\[
\hat{\theta}_0 = \big( \sum_{i\in U_s} \frac{w_i}{\sigma_i^2} u_i u_i^{\top} \big)^{-1} \sum_{i\in U_s} \frac{w_i}{\sigma_i^2} u_i y_i  
\]

Next, for an innovative approach to the RNF \eqref{RNF}, we propose 
\begin{equation} \label{tildeD}
D(\mu_U, y_U) = \sum_{i\in U} \frac{\tilde{e}_i^2 }{2 \sigma_{i+}^2}
\end{equation}
where $\tilde{e}_i$ is given via the same neighbour function \eqref{lineardot} for $\dot{\mu}_i$ as
\[
\tilde{e}_i = e_i - \lambda \dot{e}_i = e_i -  \lambda \sum_{j\in \nu_i} m_{ij} e_j = \tilde{y}_i - c_i
\quad\text{with}\quad \tilde{y}_i = y_i - \lambda \dot{y}_i
\]
and 
\[
\sigma_{i+}^2 = \sigma_i^2 + \lambda^2 \sum_{j\in \nu_i} m_{ij}^2 \sigma_j^2
\]
Note that $\sigma_{i+}^2$ would be the variance of $\tilde{e}_i$ if $e_i$ is a random variable with variance $\sigma_i^2$ and is independent across the nodes. The SEE \eqref{SEE} follows, where
\[
\begin{cases} 
H_i(\lambda) = - \sigma_{i+}^{-2} \tilde{e}_i \dot{y}_i - \lambda \sigma_{i+}^{-4} \tilde{e}_i^2 \sum_{j\in \nu_i} m_{ij}^2 \sigma_j^2 \\
H_i(\beta) = \sigma_{i+}^{-2} c'(x_i; \beta) (c_i - \tilde{y}_i) \quad\text{with}\quad c'(x_i; \beta) = \partial c(x_i ; \beta) / \partial \beta
\end{cases}
\]

As a practical alternative, one could choose a grid of values $\lambda$ subject to the range restriction of $\lambda_0$, find the corresponding $\beta_{\lambda}$ for each $\lambda$ by solving
\[
H(\beta; G_s \mid \lambda) = \sum_{i\in s} \frac{w_i}{\sigma_{i+}^2} c'(x_i; \beta) (c_i - \tilde{y}_i) = 0
\]
and choose as $\hat{\theta}_0$ the pair of $(\lambda, \beta_{\lambda})$ that minimise the estimate of \eqref{tildeD} given by
\[
\hat{D}(\mu_U, y_U) = \sum_{i\in s} \frac{w_i \tilde{e}_i^2 }{2 \sigma_{i+}^2}
\]

\subsection{Graph sampling and weighting} \label{sampling}

The way by which the edges drive graph sampling is called the \emph{observation procedure (OP)}; see Zhang (2022), Zhang and Patone (2017). In particular, we assume in this paper that the OP is \emph{reciprocal incident} such that, given an initial sample of nodes $s_0$, both $a_{ij}$ and $a_{ji}$ are observed for any $i\in s_0$. For instance, if $s_0 = \{ i_4\}$ in Figure \ref{fig:toy}, then $\{ i_2, i_3\}$ are observed as well as $a_{i_2 i_4}$ and $a_{i_3 i_4}$, i.e. the direction of an edge (in digraphs) does not matter under the reciprocal incident OP. Moreover, the values in $v_G$ are assumed to be observed together with the associated nodes and edges. Finally, an OP may be applied repeatedly, i.e. to the observed nodes outside the initial $s_0$, and the set of nodes to which the OP is applied is called the \emph{seed sample}, denoted by $s$. Let $A_s$ contain all the elements of $A$ which are observed from the seed sample. Zhang (2022) defines sample graph to be $G_s = (U_s, A_s)$, where $U_s = s \cup \{ i,j : a_{ij}\in A_s\}$. 

For any $i\in U_s$, denote by $\delta_i = 1$ or 0 whether all the values required for $H_i(\theta)$ are observed. Given \eqref{D}, we have $\delta_i =1$ for the CNF iff $(x_{\nu_i}, \omega_{\nu_i})$ are observed; given \eqref{tildeD}, we have $\delta_i =1$ for the RNF iff $(y_{\nu_i}, x_{\nu_i}, \omega_{\nu_i})$ are observed. Thus, although the RNF $\mu = Q c$ depends on the whole graph, we only need to observe the neighbourhood of $i$ and all the associated values, in order to be able to use it for learning, as long as we adopt the proposed metric \eqref{tildeD} for RNF. In contrast, for the $Q_{\tau}$-approximation $\mu = Q_{\tau} c$, we have $\delta_i =1$ iff $(y_{\nu_i^{\tau}}, x_{\nu_i^{\tau}}, \omega_{\nu_i^{\tau}})$ are observed, where $\nu_i^{\tau}$ is defined by \eqref{ext} and $\omega_{\nu_i^{\tau}}$ are associated with all the paths starting from $i$ and consisting of at most  $\tau$ edges (regardless their directions). 
The observation requirement is actually more demanding than for the RNF.

Below we describe the details of sampling and weighting for NF learning. By virtue of the reciprocal incident OP, the description given below in terms of undirected graphs applies equally to digraphs.

\subsubsection{Targeted random walk (TRW)}  

There are many types of random walk on graphs; see e.g. Masuda et al. (2017). We consider the following.  
Starting from any node, denoted by $O_0 = i \in U$, we either move randomly over one of the edges incident to $i$ or jump randomly to any node by the following transition probabilities (Avrachenkov, et al., 2010)
\[
p_{ij} = \begin{cases} \frac{1}{d_i + r} \big( 1 + \frac{r}{N} \big) & \text{if } a_{ij} =1 \\ 
\frac{r}{d_i +r} \big( \frac{1}{N} \big) & \text{if } a_{ij} = 0 \text{ including } i=j \end{cases}
\]
where $r$ is a chosen tuning constant. This yields $O_1 = j$, and so on. By virtue of random jump, the Markov process $\{ O_t : t\geq 0 \}$ is irreducible. On reaching its equilibrium, the stationary probability is given by
\begin{equation} \label{TRW}
\pi_i = \Pr(O_t = i) \propto d_i + r
\end{equation} 
The same stationary probabilities $\{ \pi_i : i\in U\}$ can as well be obtained by faster-mixing lagged random walk (Zhang, 2021).

Applying the OP to the current state $O_t = i$ enables one to observe $\nu_i$ and the associated $(x_{\nu_i}, y_{\nu_i}, \omega_{\nu_i})$. Hence, for both the CNF \eqref{CNF} and RNF \eqref{RNF}, we have $\delta_i = 1$ iff $i$ is in the seed sample of TRW. Given an extraction of $n$ states at equilibrium, denoted by $s_n = \{ t_1, ..., t_n\}$, the SEE \eqref{SEE} can be written as
\[
H(\theta; G_s) = \sum_{k=1}^n \sum_{i\in U} \mathbb{I}(O_{t_k} =i) w_i H_i(\theta) 
\]
For each $i \in U$, setting
\[
w_i^{-1} = n (d_i + r)
\]
makes the SEE is unbiased, since
\[
E\big[ H(\theta; G_s) \big] = \frac{1}{n} \sum_{k=1}^n \sum_{i\in U} \frac{\pi_i}{d_i + r} H_i(\theta) \propto \sum_{i\in U} H_i(\theta) = 0
\]

Under TRW, the states are correlated in the seed sample $s_n = \{ t_1, ..., t_n\}$. The simplest approach to unbiased variance estimation is to run multiple TRWs independently and obtain, say, $\hat{\theta}_l$ for $l =1, ..., L$. The combined estimator and variance estimator are then given by
\[
\hat{\theta} = \frac{1}{L} \sum_{l=1}^L \hat{\theta}_l \qquad\text{and}\qquad 
\hat{V}(\hat{\theta}) = \frac{1}{L(L-1)} \sum_{l=1}^L (\hat{\theta}_l - \hat{\theta})^2 
\]

\subsubsection{Snowball sampling (SBS)} 

Let $s_0\sim p(s_0)$ for the initial sample $s_0 \subset U$. For $t=1, ..., T$, let
\[
s_t = \nu(s_{t-1}) \setminus \bigcup_{r=0}^{t-1} s_r 
\]
be the $t$-th wave sample of nodes, which are outside of $s_{t-1}$ but adjacent to some nodes in $s_{t-1}$. The sampling is terminated if $s_t = \emptyset$ for some $t< T$, in which case $s_r = \emptyset$ for $r = t, ..., T$. The seed sample of \emph{$T$-wave snowball sampling ($T$SBS}) is 
\[
s =\bigcup_{t=0}^{T-1} s_t
\]
(Zhang, 2022; Frank and Snijders, 1994; Goodman, 1961). By the reciprocal incident OP, the elements of $A_s$ are $s\times U \cup U\times s$.

Clearly, 1SBS is feasible for the learning of CNF \eqref{CNF} or RNF \eqref{RNF}, where $\delta_i = 1$ iff $i\in s_0$. The number of nodes with $\delta_i =1$ can be increased by $T$SBS, if $T > 1$. For example, suppose $s_0 = \{ i_2\}$ in Figure \ref{fig:toy}. By 1SBS, we observe all the other nodes, $s_1 = \{ i_1, i_3, i_4\}$, but not yet \emph{all} the neighbours of these nodes because it is unclear whether there are any edges among $s_1$. After the 2nd wave, all the neighbours of $s_1$ are also observed, so that $\delta_i = 1$ for $s_0 \cup s_1$. Thus, by $T$SBS generally, we have $\delta_i =1$ iff $i$ is in the seed sample $s$. 

To define the weights, we use the \emph{sample-dependent} strategy (Zhang, 2022). Given the sample graph $G_s = (U_s, A_s)$, let $F_{i,s}$ contain all the nodes (including itself) that can lead to $i\in s$ (i.e. $\delta_i =1$) by $T$-wave OP \emph{in} $G_s$, in contrast to $F_i$ that contains all such nodes in $G$. Notice that $F_{i,s}$ is generally not equal to $F_i$ because $G_s \neq G$. For the SEE \eqref{SEE}, let $w_i =0$ if $i\in U_s\setminus s$ and, for any $i\in s$, let 
\begin{equation} \label{TSBS}
w_i^{-1} = \Pr(s_0\cap F_{i,s} \neq \emptyset) = 1- \Pr(s_0\cap F_{i,s} = \emptyset)
\end{equation}
according to the sampling design of $s_0$. The SEE is then unbiased \emph{conditional} on the chosen $F_{i,s}$, which means that we let $\delta_i =1$ \emph{only if} at least one node in $F_{i,s}$ is selected in the initial sample but not otherwise, i.e.
\[
E\big[ H(\theta; G_s) \big] = \sum_{i\in U} E(\delta_i) w_i H_i(\theta) 
= \sum_{i\in U} \Pr(s_0\cap F_{i,s} \neq \emptyset) w_i H_i(\theta) =  \sum_{i\in U} H_i(\theta) 
\]

\begin{figure}[ht] 
\centering
\begin{tikzcd}[cramped]
i_1 \arrow[r] & i_2 \arrow[r] & i_3 \arrow[r] \arrow[l] & i_4 \arrow[r] & i_5 \arrow[l] 
\end{tikzcd} 
\caption{A digraph with 5 nodes} \label{fig:line}
\end{figure}
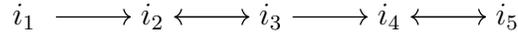

That is, under hypothetically repeated sampling, we would set $\delta_i =0$ despite $i\in s$, whenever $i\in s$ happens only because $(F_i \setminus F_{i,s}) \cap s_0 \neq \emptyset$. Take the digraph in Figure \ref{fig:line}. Suppose 3SBS yields $s_0 = \{ i_1 \}$, $s_1 = \{ 2\}$, $s_2 = \{ i_3 \}$ and $s_3 = \{ i_4\}$, such that $F_{i_3, s} = \{ i_1, i_2, i_3, i_4\}$ for $i_3 \in s$, although $i_5 \in F_{i_3}$ as well. By the strategy \eqref{TSBS}, we would set $\delta_{i_3} =0$ whenever 3SBS starts with $s_0 = \{ i_5 \}$ under repeated sammpling, even though $i_3$ is included in the seed sample $s$ then. 

Let $\hat{\theta}$ be the solution to SEE. By expansion around $\theta_0$, we have
\[
0 = \sum_{i\in U_s} w_i H_i(\hat{\theta}) \approx \sum_{i\in U} \delta_i w_i H_i(\theta_0) 
+ \Big[ \sum_{i\in U} H_i'(\theta_0) \Big] (\hat{\theta} - \theta_0)
\] 
where $H_i'(\theta_0) = \partial H_i(\theta_0) / \partial \theta_0$. The sampling variance of $\hat{\theta}$ follows as
\begin{equation} \label{var}
V(\hat{\theta}) \approx \Big[ \sum_{i\in U} H_i'(\theta_0) \Big]^{-1} 
\Big[ \sum_{i,j\in U} \Delta_{ij} H_i(\theta_0) H_j(\theta_0)^{\top} \Big]
\Big[ \sum_{i\in U} H_i'(\theta_0) \Big]^{-1}
\end{equation}
where 
\[
\Delta_{ij} = w_i w_j \Pr( \delta_i \delta_j =1) - 1 
\]
and
\[
\Pr( \delta_i \delta_j =1) = 1 - \Pr( s_0 \cap F_{i,s} = \emptyset) - \Pr( s_0 \cap F_{j,s} = \emptyset) 
+ \Pr\big( s_0 \cap (F_{i,s} \cup F_{j,s}) = \emptyset \big)
\]

\paragraph{Remark} The strategy consisting of $T$SBS and \eqref{TSBS} is also applicable to the learning of $\mu = Q_{\tau} c$. Let $s_{\tau}$ be the subset of seed sample $s$, containing all the nodes for which $\nu_i^{\tau}$ defined by \eqref{ext} is observed under $T$SBS. Note that $s_0 \subseteq s_{\tau}$ provided $T\geq \tau$, since $\nu_i^{\tau}$ is always observed for any $i$ in $s_0$ under $\tau$SBS. Let $F_{i,s}$ and the weight \eqref{TSBS} be as defined above for any node $i\in s_{\tau}$.  
 
For an illustration, suppose $(T, \tau) = (3,2)$, where 3SBS starts from $s_0 = \{ i_1 \}$ in Figure \ref{fig:line}. We have $i_3 \not\in s_{\tau} = \{ i_1, i_2\}$ because it is unclear if $\nu_{i_3}^2$ is observed, now that $i_4$ is observed on the last wave and it is possible that $i_4$ is adjacent to other yet unobserved nodes, which indeed is the case here. The weight \eqref{TSBS} can be calculated for $F_{i_1,s} = \{ i_1,  i_2, i_3 \}$ and $F_{i_2,s} = \{ i_1,  i_2, i_3, i_4 \}$.

Clearly, the graph computation required for the learning of $\mu = Q_{\tau} c$ is more complicated without ready-made software.


\begin{thebibliography}{999}

\bibitem{avrachenkov2010} Avrachenkov, K., Ribeiro, B. and Towsley, D. (2010). Improving Random Walk Estimation Accuracy with Uniform Restarts. \emph{Research report}, RR-7394, INRIA. inria-00520350 

\bibitem{frank1994estimating} Frank O. and Snijders T. (1994). Estimating the size of hidden populations using snowball sampling. {\em Journal of Official Statistics}, 10:53-53.

\bibitem{friedkin1990} Friedkin, N.E. (1990). Social Networks in Structural Equation Models. \textit{Social Psychology Quarterly}, 53:316-328.

\bibitem{goodman1961} Goodman, L.A. (1961). Snowball sampling. \textit{Annals of Mathematical Statistics}, 32:148-170.

\bibitem{leenders2002} Leenders, R. (2002). Modeling social influence through network autocorrelation: constructing the weight matrix. \textit{Social Networks}, 24:21-47.

\bibitem{masuda2017} Masuda, N., Porter, M.A. and Lambiotte, R. (2017) Random walks and diffusion on networks. \emph{Physics Reports}, 716-717: 1-58. 

\bibitem{ord1975} Ord, K. (1975). Estimation Methods for Models of Spatial Interaction. \textit{Journal of the American Statistical Association}, 70:120-126.

\bibitem{lcz2022} Zhang, L.-C. (2022). \emph{Graph sampling.} CRC Press. 

\bibitem{lcz2021} Zhang, L.-C. (2021a). Graph sampling by lagged random walks. \textit{Stat}, \url{https://onlinelibrary.wiley.com/doi/abs/10.1002/sta4.444}

\bibitem{zhang2017graph} Zhang, L.-C. and Patone, M. (2017). Graph sampling. \emph{Metron}, 75:277-299.

\end{thebibliography}
\end{document}